# VIS: the visible imager for *Euclid*

Mark Cropper*[a], S. Pottinger[a], S. Niemi[a], R. Azzollini[a], J. Denniston[a], M. Szafraniec[a], S. Awan[a], Y. Mellier[b,c], M. Berthe[c], J. Martignac[c], C. Cara[c], A.-M. di Giorgio[d], A. Sciortino[e], E. Bozzo[f], L. Genolet[f], R. Cole[a], A. Philippon[g], M. Hailey[a], T. Hunt[a], I. Swindells[h], A. Holland[i], J. Gow[i], N. Murray[i], D. Hall[i], J. Skottfelt[i], J. Amiaux[c], R. Laureijs[j], G. Racca[j], J.-C. Salvignol[j], A. Short[j], J. Lorenzo Alvarez[j], T. Kitching[a], H. Hoekstra[k], R. Massey[l], H. Israel[l,m] and the *Euclid* collaboration

[a]Mullard Space Science Laboratory, University College London, Holmbury St Mary, Dorking Surrey RH5 6NT, United Kingdom; [b]Institut d'Astrophysique de Paris, 98 bis Boulevard Arago, 75014 Paris, France; [c]Service d'Astrophysique, Commissariat à l'Énergie Atomique, Orme des Merisiers, Bat 709, 91191 Gif sur Yvette, France; [d]Istituto di Astrofisica e Planetologia Spaziali, INAF, Via del Fosso del Cavaliere, 100, 00133 Roma, Italy; [e]CGS S.p.A., Milan, Italy; [f]ISDC Data Centre for Astrophysics, Chemin d'Ecogia 16, CH-1290 Versoix, Switzerland; [g]Institut d'Astrophysique Spatiale, Campus Universitaire d'Orsay, Batiment 121, Orsay cedex 91405, France; [h]e2v technologies plc, 106 Waterhouse Lane, Chelmsford, Essex CM1 2QU, United Kingdom; [i]Centre for Electronic Imaging, Planetary and Space Sciences Research Institute, The Open University, Walton Hall, Milton Keynes, MK7 6AA, United Kingdom; [j]European Space Agency / ESTEC, Keplerlaan 1, 2201 AZ Noordwijk, The Netherlands; [k]Leiden Observatory, Leiden University, Niels Bohrweg 2, NL-2333 CA, Leiden, The Netherlands; [l]Department of Physics, Durham University, South Road, Durham DH1 3LE, UK; [m]Faculty of Physics, Ludwig-Maximilians Universität, Scheinerstr. 1, 81679 Munich, Germany.

## ABSTRACT

*Euclid*-VIS is the large format visible imager for the ESA *Euclid* space mission in their Cosmic Vision program, scheduled for launch in 2020. Together with the near infrared imaging within the NISP instrument, it forms the basis of the weak lensing measurements of *Euclid*. VIS will image in a single $r+i+z$ band from 550-900 nm over a field of view of ~0.5 deg$^2$. By combining 4 exposures with a total of 2260 sec, VIS will reach to deeper than $m_{AB}$=24.5 (10σ) for sources with extent ~0.3 arcsec. The image sampling is 0.1 arcsec. VIS will provide deep imaging with a tightly controlled and stable point spread function (PSF) over a wide survey area of 15000 deg$^2$ to measure the cosmic shear from nearly 1.5 billion galaxies to high levels of accuracy, from which the cosmological parameters will be measured. In addition, VIS will also provide a legacy dataset with an unprecedented combination of spatial resolution, depth and area covering most of the extra-Galactic sky. Here we will present the results of the study carried out by the *Euclid* Consortium during the period up to the Critical Design Review.

**Keywords:** dark energy, dark matter, cosmology, visible light imaging.

## 1. INTRODUCTION

In the ΛCDM model of the Universe, approximately three quarters consists of dark energy, and approximately one fifth of dark matter. The nature of these constituents is largely unknown. *Euclid*, the second mission in ESA's Cosmic Vision programme, is designed to make accurate measurements to infer the nature of dark energy and to quantify its role in the evolution of the Universe. *Euclid* will additionally map and elucidate the nature of dark matter. If, instead, the dark energy is a manifestation of a deviation from general relativity on cosmic scales, then *Euclid* will also test the validity of many of these modified gravity theories. Until there is high accuracy data to put these new theoretical frameworks to the test, real progress in constraining the nature of the Cosmos will be limited. *Euclid* will be one of the most powerful tools in this quest, one in which the systematics will be controlled to an unprecedented accuracy through the combination of technical capability, data calibration and simulation and by employing different cosmological approaches.

---

* m.cropper@ucl.ac.uk; www.ucl.ac.uk/mssl

Besides these studies in physics and cosmology, *Euclid* will provide a colossal legacy dataset over the whole sky, with optical imaging at 0.2 arcsec spatial resolution to very faint limits, $m_{AB}\sim24.5$ at $10\sigma$ in a broad $r+i+z$ band, infrared imaging in three bands, and spectra and redshifts of millions of galaxies out to redshift 2. A dataset of this size will be used by scientists worldwide in a wide range of contexts, and it will have huge scientific and public impact.

This paper discusses the Visible Imager (VIS) on *Euclid*, as it enters into its flight-engineering phase (mid-2016), updating the position of earlier phases[1,2,3,4]. VIS is complemented by the Near Infrared Spectrometer Photometer (NISP) described in a companion paper[5]. The overall scientific aims of *Euclid*, and an overview of the mission are available in the *Euclid* Red book[6] and in these proceedings[7].

## 2. PERFORMANCE REQUIREMENTS

### 2.1 Science aims

The core task of VIS is to enable Weak Lensing measurements[8,9]. The dark matter (and ordinary matter) aggregates under the influence of gravity as the Universe expands. These overdensities deflect light differentially so that the light from background objects is distorted, and they appear to have a measurable change in ellipticity. In general there is only mild distortion, corresponding to weak gravitational lensing. The mass distribution can be inferred from statistical averages of the shapes of background galaxies distorted by this effect, enabling *Euclid* to map dark matter. By using galaxies further and further away, the rate at which the aggregation has occurred can be characterised. This is directly affected by the expansion history of the Universe, which appears at more recent times to be increasingly driven by dark energy, so the characteristics of the dark energy can consequently also be constrained.

While there is a chain of inferences, Weak Lensing is considered to be one of the most powerful techniques to determine the characteristics of dark matter and dark energy[10,11]. To accomplish this task requires very large surveys in order to ensure a sufficient number of sources and to overcome the different apparent galaxy shapes and natural variations within the Universe. Also required are extremely accurate measurements of the galaxy shapes. Systematic effects must be deeply understood and a prerequisite for this is calibrations of the highest quality and an advanced approach in modelling the effects. The scientific performance depends largely on the difference between this understanding, embodied in the model, and the unknown truth within the measurements.

### 2.2 VIS characteristics

Therefore, VIS requires a large field of view sampled sufficiently finely to measure typical galaxy shapes. To cover most of the extra-Galactic sky in a reasonable mission duration (5–6 years), with a telescope commensurate with a Medium mission in the Cosmic Vision programme (which dictates the exposure duration to meet the limiting magnitude) the field of view must be $\sim0.5$ deg$^2$. To sample galaxies whose typical sizes are $\sim0.3$ arcsec, pixel sizes of 0.1 arcsec and smaller are required. To minimise the mass of the focal plane and the payload as a whole, the image scale must not be too large, so detector pixels must be as small, yet consistent with a sufficient dynamic range (otherwise even faint objects will saturate) and within a proven technological capability. These requirements are met with a VIS focal plane of 36 CCDs, each of 4kx4k pixels, each 12μm square. Only CCDs have the requisite performance characteristics or track record in space. With 604 Mpix, VIS will have the second largest focal plane (after ESA's *Gaia* satellite) to be flown in space. Unlike *Gaia*, all of the image data from the focal plane will be telemetered to ground.

Combined with the NISP infrared photometric measurements and data from the ground, the Weak Lensing measurements do not require VIS to provide multicolour information within the optical band[12,13]. Shear measurements of distant galaxies at blue wavelengths suffer more from the more clumpy appearance of galaxies in the ultraviolet (shifted by the expansion of the Universe into the visible band). VIS therefore implements a single broad red band.

Beyond these broad top-level requirements, VIS requires a shutter, a calibration unit for flat-fielding the detector, and electronics units to process the data from the large focal plane and to control the instrument.

The more general requirements for VIS are given in Table 1.

Table 1: VIS and weak lensing channel characteristics

| Spectral Band | 550 – 900 nm |
|---|---|
| System Point Spread Function size | ≤0.18 arcsec full width half maximum at 800 nm |
| System PSF ellipticity | ≤15% using a quadrupole definition |
| Field of View | >0.5 deg$^2$ |
| CCD pixel sampling | 0.1 arcsec |
| Detector cosmetics including cosmic rays | ≤3% of bad pixels per exposure |
| Linearity post calibration | ≤0.01% |
| Distortion post calibration | ≤0.005% on a scale of 4 arcmin |
| Sensitivity | $m_{AB}$≥24.5 at 10σ in 3 exposures for galaxy size 0.3 arcsec |
| Straylight | ≤20% of the Zodiacal light background at Ecliptic Poles |
| Survey area | 15000 deg$^2$ over a nominal mission with 85% efficiency |
| Mission duration | 6 years including commissioning |
| Shear systematic bias allocation | additive $\sigma_{sys}$ ≤ 2 x 10$^{-4}$ ; multiplicative ≤ 2 x 10$^{-3}$ |

## 3. EXTERNAL INTERFACES

### 3.1 Optical interface

*Euclid* has a 1.2m 3-mirror telescope, with the third mirror providing corrections to the intermediate Cassegrain focus[14]. The VIS and NISP beams from the shared field of view are split by a dichroic, working in reflection for VIS to produce a bandpass of 550 to 900nm with ~20nm band edges (Figure 1). The folding mirrors carry multi-layer coatings to provide further bandpass control and out-of-band rejection. To minimise the brightness of ghost images from internal reflections, especially as the rays incident on the focal plane are not telecentric, there are no other filters or optics in VIS. The entrance pupil is 1.2m, the central obscuration is 13% and the final focal ratio is f/20.4. This provides an image scale of 0.101 arcsec per 12μm pixel, and a field of view of 0.72°x0.79°, or just under 0.57 square degrees. The optical performance is diffraction limited at 800 nm over the entire field; however, given the tight requirements on the PSF knowledge for weak lensing measurements, the PSF nevertheless is variable over the focal plane and, as a function of time, dependant on the thermo-elastic stability of the optical system (section 7.2 below).

### 3.2 Thermal and mechanical

VIS has five assemblies (Section 4). Three of the five VIS assemblies – the focal plane, the shutter and the calibration unit – are supported by the Payload Module structure in the space below the telescope with separate interfaces to the structure. These are shown in Figure 1. The other two modules are located in the warm Service Module. Besides the normal requirements for withstanding launch stresses, the positioning of the focal plane with respect to the telescope focus is the most critical mechanical interface.

A cold environment ~150K is provided in the *Euclid* Payload Module to facilitate the operation of the VIS CCDs, which show optimal charge transfer efficiency at this temperature (see section 7.1). On the other hand, for reasons of device operating conditions, the CCD detection chain electronics must be maintained at a temperature of ≥250K. To provide the two different temperature regimes, a radiator to space is provided by the spacecraft for the detection chain electronics while the detector plane is cooled by radiation into the cold Payload Module.

The total mass of VIS is 104 kg including 10% margin. This mass is relatively stable, 19 kg within allocation.

### 3.3 Electrical interfaces

The spacecraft Command and Data Management Unit (CDMU) controls VIS through a MIL-STD-1553 bus. Data are transferred to the on-board bulk memory MMU through high speed SpaceWire links. Primary power (28V, unregulated) is provided by the spacecraft Power Control System (PCDU).

VIS uses 201W of primary power (214W peak) including 10% margin, 52W within allocation. It produces 423 Gbit of data per day (compressed without loss by a factor 2.8), 23% within margin.

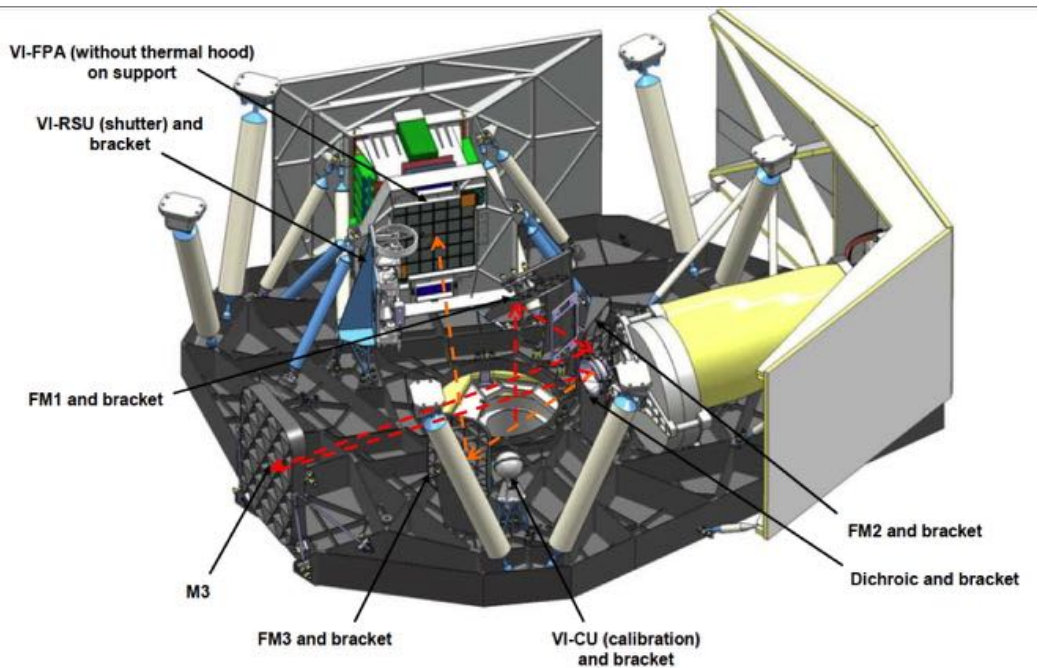

Figure 1: The placement of the VIS instrument within the Payload Module (VI-FPA, VI-RSU, VI-CU – see Figure 2 for details). The large radiator for the VIS detection chain electronics (the ROE) is on the far side. The yellow instrument is the NISP. The PLM is shown "upside down" with the telescope below it. The light path from the hole in the primary mirror is shown in red before the dichroic, and orange after it. FMs are fold mirrors, and M3 is the third mirror in the Korsch telescope. Figure courtesy of Airbus Defence & Space.

## 4. DESIGN

The five VIS assemblies are shown in Figure 2, with an expanded view of the Focal Plane Array in Figure 3. The electrical architecture is shown in Figure 4.

### 4.1 Focal Plane Array

Photons within the broad red bandpass (550–900 nm) are detected by the array of 36 CCDs, each of which is a CCD273-84[15,16] specifically designed for VIS and manufactured by e2v Technologies. The CCDs are located in a 6x6 matrix on the front of the Focal Plane Array (FPA). Close butting of the CCDs provides a >90% filling fraction of active Si.

The FPA has dimensions of ~0.45m on each side. The 36 CCDs are held in a Silicon Carbide (SiC) structure[17,18]. The detector array is maintained relative to the optical focal plane with tight tolerances to ensure image quality. There is a separate block of electronics which digitise the signals from the CCDs, and is supported separately, with only the CCD electrical flexible interconnects between them (Figure 3). The electronics block holds twelve Readout Electronics (ROE – one for three CCDs), and their closely coupled power supply units (R-PSU, one per ROE). Within the support structure for the ROEs are thermal shields to isolate the cold CCDs from the warm ROEs. A radiator to space maintains the ROE thermal environment.The detector block with the 36 CCDs is maintained at a temperature of 150–155K by the PLM environment.

The detection chain consists of the CCDs, ROEs and their associated R-PSUs. The architecture of the chain has been set up to ensure sufficient redundancy, while minimising the use of system resources. Manufacturability, testability and ease of access were important considerations. The tradeoff identified the optimal configuration as one ROE supporting three CCDs, each with four video chains (Figure 5), each with its own power supply unit. In order to minimise system noise levels, the power supply for each ROE is accommodated close to it on the outside of the Electronics Block. A single clocking and bias generation circuit is implemented for each CCD half. Loss of one ROE results in loss of <10% of the focal plane, and the consequent thermal perturbations are manageable. Within each ROE it is envisaged that individual CCDs may be lost without loss of the entire unit, providing further redundancy.

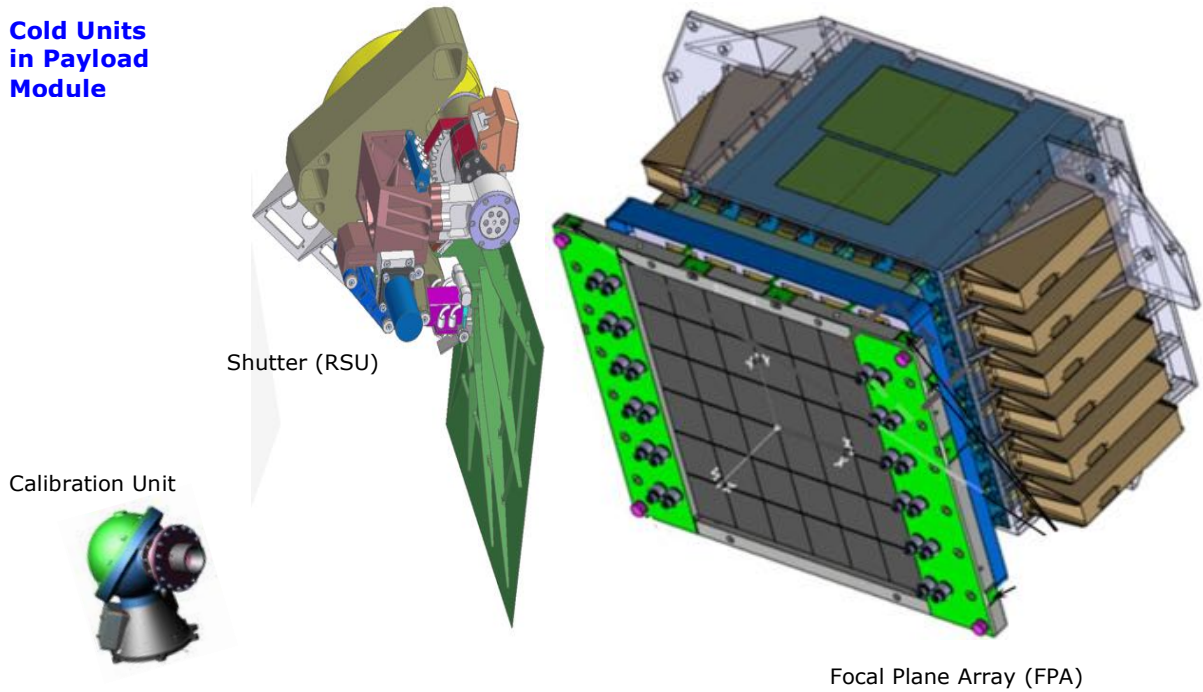

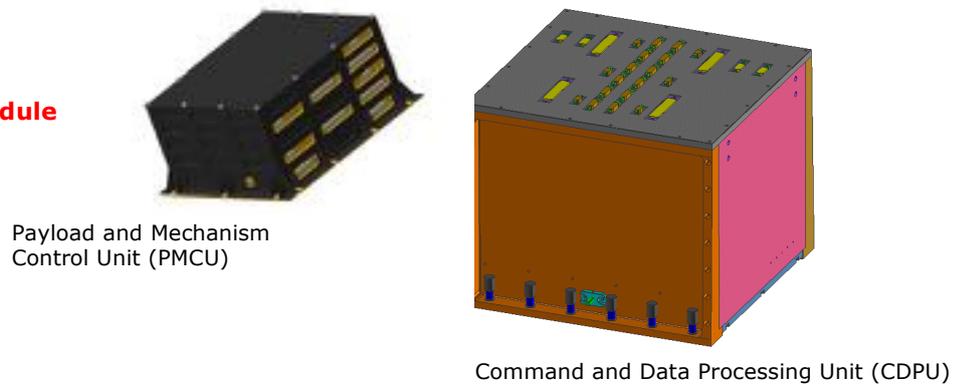

Figure 2: The five units comprising VIS. The two units at the bottom are in the Service Module and the other three in the Payload Module.

The CCDs are read out through four readout nodes at a rate of 70 kpix s$^{-1}$. Analogue signals are converted to 16 bits resolution. In order to maintain thermal stability, all circuitry remains powered during exposures. Care is taken in the design of mixed signal circuitry handling very low level analogue inputs, to prevent cross-talk and other noise pick-up: multi-layer printed circuit boards are used with separate ground planes for analogue and digital functions. System grounding and decoupling is carefully planned to prevent circulating currents in ground lines from introducing noise sources. All of the CCDs are read out synchronously, removing electrical co-interference that could result from a slew of non-synchronised clocks; an LVDS (low-voltage differential signaling) interface (Figure 4) is used for the synchronisation.

Data are transmitted through a SpaceWire port to the payload data handling unit (Figure 4), with SpaceWire communication and command decoding, together with other digital functions such as clock sequence generation carried out in a single space-qualified field-programmable gate array (FPGA) per ROE. The packaging of the ROE and PSU electronics is designed to optimize conductive thermal paths while minimising the parasitic heat conduction to the CCDs.

## 4.2 Shutter

A shutter is located in front of the FPA to block the light to the detector array while the detectors are not making exposures or reading them out. This is a momentum-compensated (linear and angular) mechanism in order to minimise any disturbances to the spacecraft and the NISP instrument during actuation. It opens and closes within 10 seconds. The mechanism is electrically cold-redundant. The previously envisaged launch lock is now considered unnecessary and, pending positive results from testing, will be removed. The shutter does not seal against any Payload Module structure, and therefore scattered light paths require careful analysis. Surface treatments for the shutter minimise scattered light when it is opening and closing.

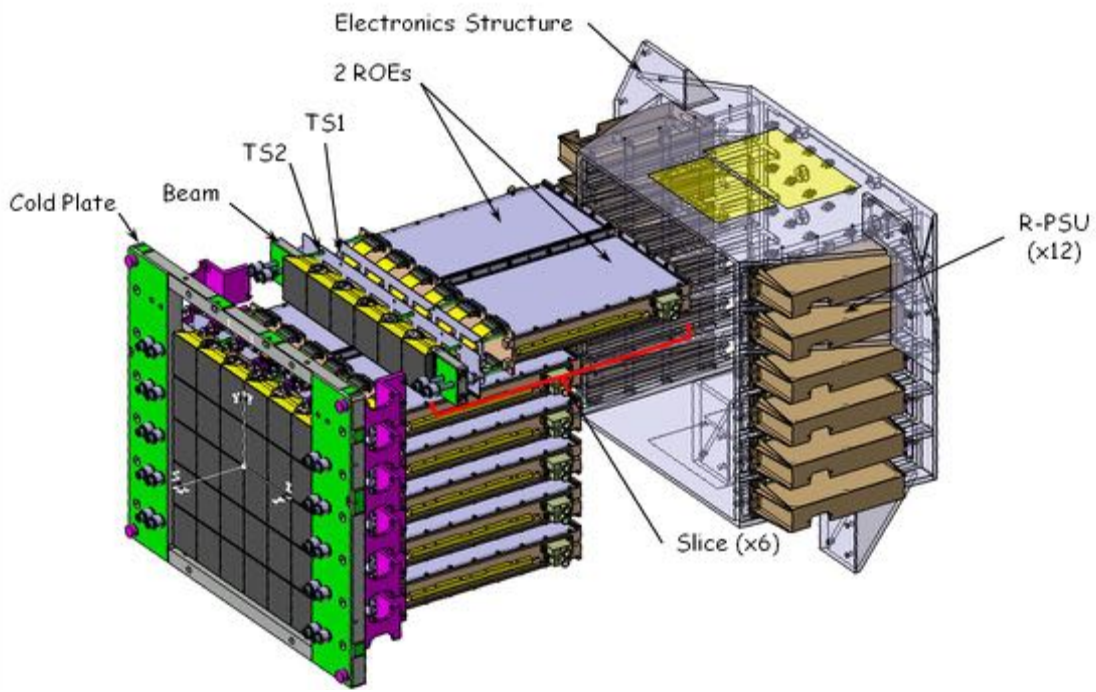

Figure 3: An expanded view of the focal plane array. The detector matrix is on the left, supported on six cold SiC beams each holding six CCDs within a SiC frame. The top row is offset to the rear, to show its relationship of each triplet of detectors to their readout electronics (ROE) which digitise the signals. The cold frame holding the CCDs is mechanically decoupled from the framework holding the ROEs. Each ROE has its own power supply unit (R-PSU). Items labelled TS are the thermal shrouds to isolate the cold detector plate from the warm ROEs.

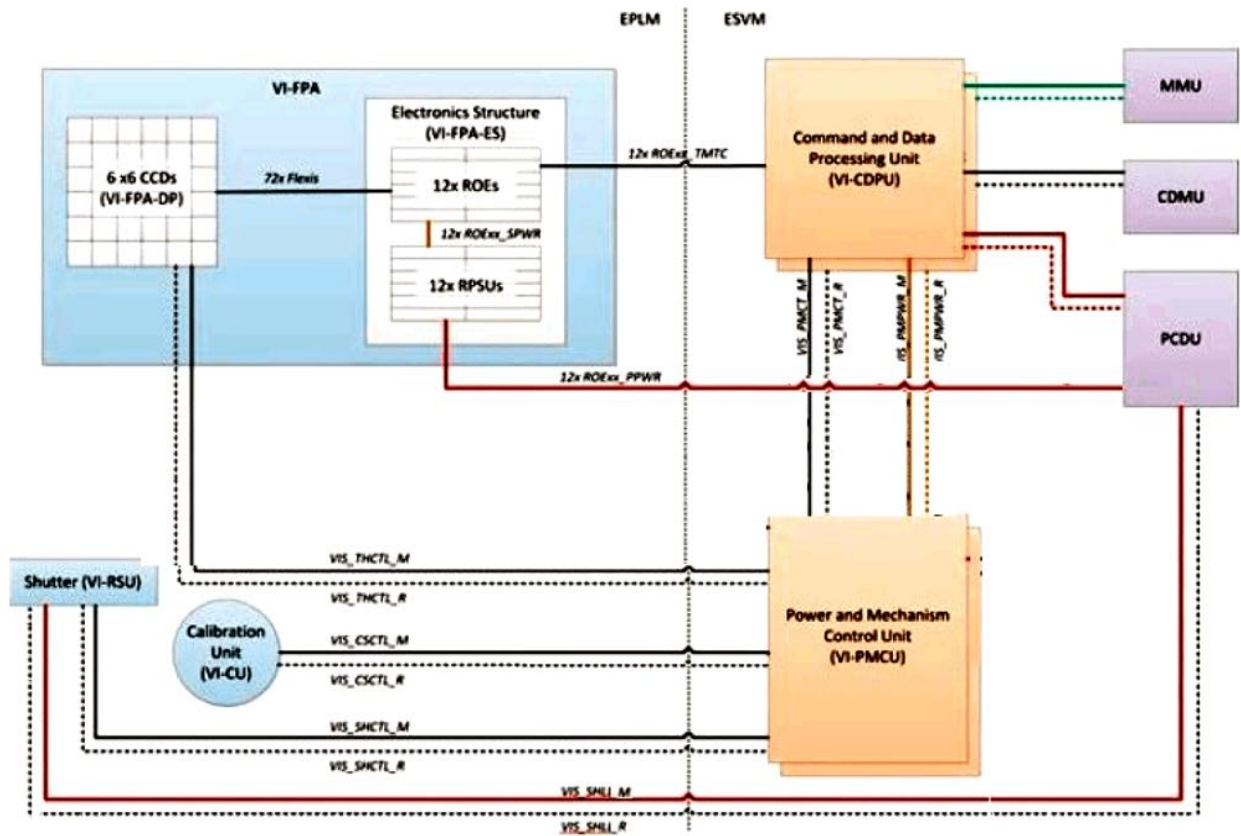

Figure 4: VIS electrical architecture. The three units to the right (MMU, CDMU and PCDU) are spacecraft units.

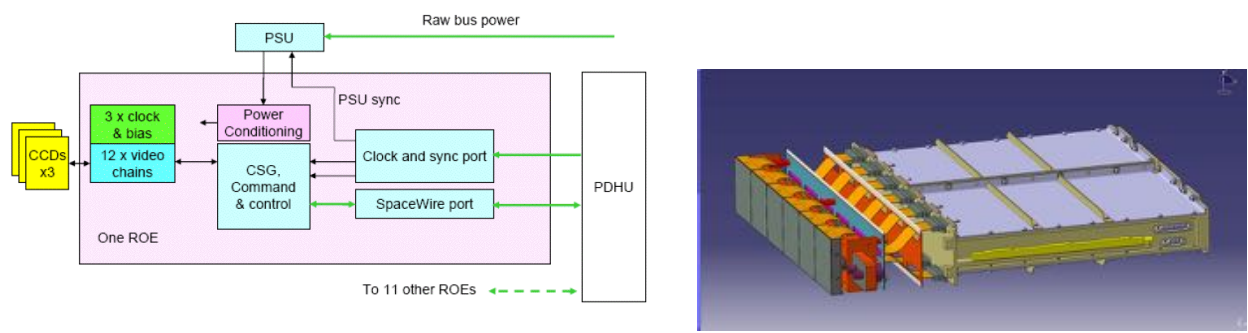

Figure 5: (left) The block diagram for an ROE, interfacing to 3 CCDs. (right) The mechanical packaging of two ROEs each with 3 CCDs to make a slice for 6 CCDs.

### 4.3 Calibration Unit

A calibration unit is provided to flood the FPA with light at different wavelengths within the VIS passband. CCD flat fields taken with this unit allow pixel-pixel variation to be measured as a function of wavelength. It is not required to provide high levels of stability or of large-scale spatial uniformity, as the photometric throughput is calibrated from stellar sources during routine observations, so a simple lens projection within the baffle provides sufficient uniformity and control of out-of-field light. The smoothest flat fields are at ~750nm. While the wavelength dependence of the CCD flat field is slow, the number of independent illumination sources has been increased from 3 to 6 to ensure its adequate characterisation.

There is no shutter for the calibration unit. Flat field calibrations therefore have the sky signal included, but because they are short exposures, and because many flat fields are combined to produce a master flat, this simple solution is acceptable. The calibration unit is electrically cold-redundant.

### 4.4 Command and Data Processing Unit

The CDPU[19,20] in the Service Module controls the instrument, transitioning between the different instrument modes and sequencing the operations within exposures, and monitoring instrument status to generate the housekeeping data. It also takes the science data from the ROEs, losslessly-compressing it and buffering it before transfer to the spacecraft bulk memory (MMU). It consists of a Maxwell SCS750 PowerPC triple-voting processing unit which performs the compression and science packet generation, a data routing board to handle the instrument status and the science data from the ROEs, power routing hardware to the PMCU (see below) and a multiplexer board to interface the 12 ROEs to both halves of the unit. The CDPU is dual-redundant except for the multiplexer board (Figure 4).

### 4.5 Power and Mechanism Control Unit

The PMCU provides power to the Calibration Unit and the Shutter, and controls its opening and closing. It also accepts housekeeping data from the shutter and calibration unit, and heater power to the FPA to ensure an appropriate thermal environment. The PMCU is also dual-redundant.

## 5. CURRENT PROGRAMME STATUS

VIS is currently in phase C approaching its Critical Design Review. The main activities are therefore the development, build, assembly and testing of the Structure Thermal Model (STM) and Engineering Model (EM) of each unit separately. These undergo unit-level environmental, functional and performance testing. In the case of the FPA, the CCDs, ROEs and their R-PSUs are integrated into the FPA mechanical structure. The STM FPA contains a full 36 CCD focal plane and STM ROEs, integrated and aligned with flight handling and contamination control procedures and metrology. The two halves of the FPA (the detector and electronics blocks) are held with flight-representative mechanical fixtures prior to their integration in the STM payload model. The EM FPA is a fully-functional 1/6 focal plane with 6 CCDs read through 2 EM ROE+R-PSUs. The integrated FPA also undergoes environmental and functional testing. All of the units are in the process of delivery for instrument-level integration and tests. The principal instrument-level integration is electrical, as there is no single mechanical structure to support all of the units (see above).

Figure 6 to Figure 13 show images of the STM and EM development. These developments are also relevant for the Flight Model.

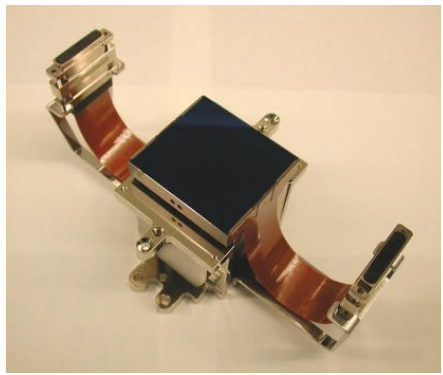

Figure 6: The CCD273-84 to be used in VIS.

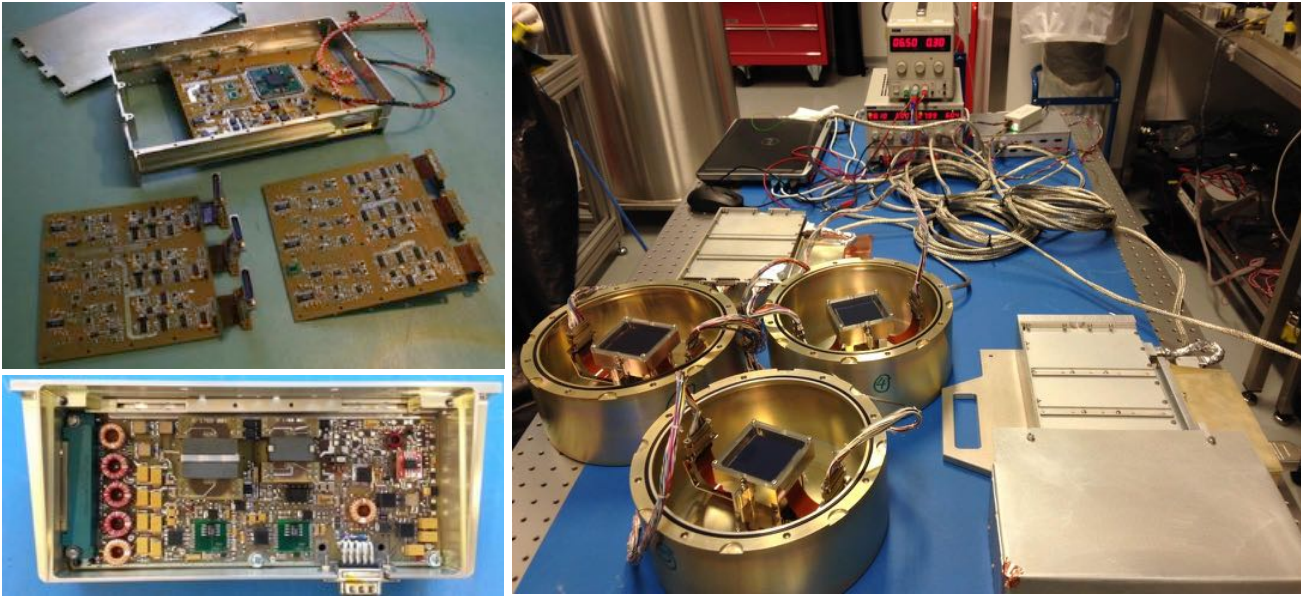

Figure 7: (top left) The EM ROE consisting of two video boards (one each for top and bottom halves of the three CCDs) and the digital board (in the housing) for clock generation and housekeeping. (bottom left) The EM R-PSU. (right) An EM ROE and R-PSU (on the lower right of the image)

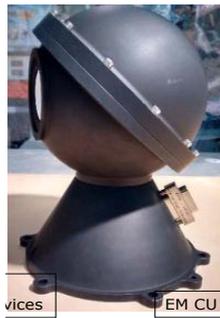

Figure 8: The EM Calibration Unit. This unit lacks the baffle on the exit port (on the left) with its field stop and lens system.

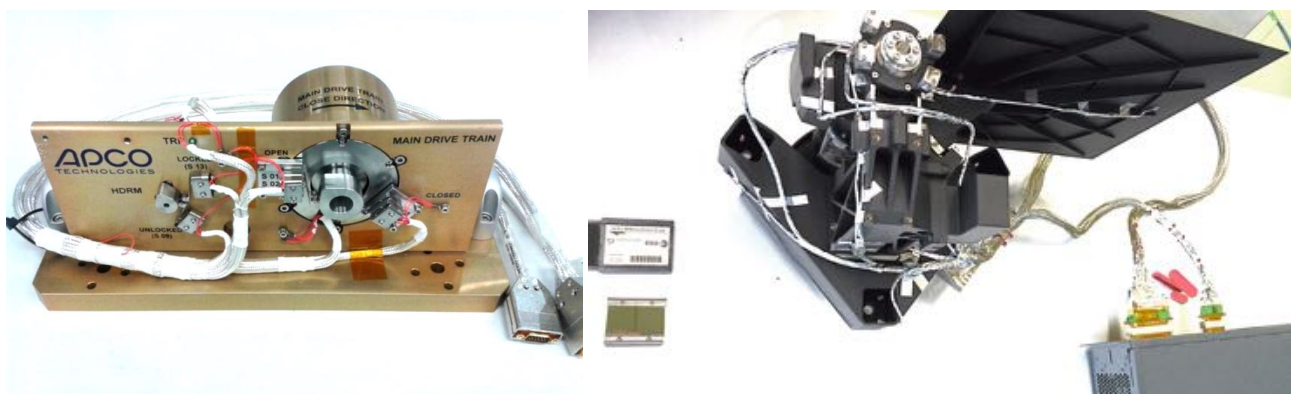

Figure 9: (left) The EM for the Shutter. (right) The STM for the shutter. The shutter leaf is at the top of the image, and covers the full CCD focal plane. The mounting structure is in the centre.

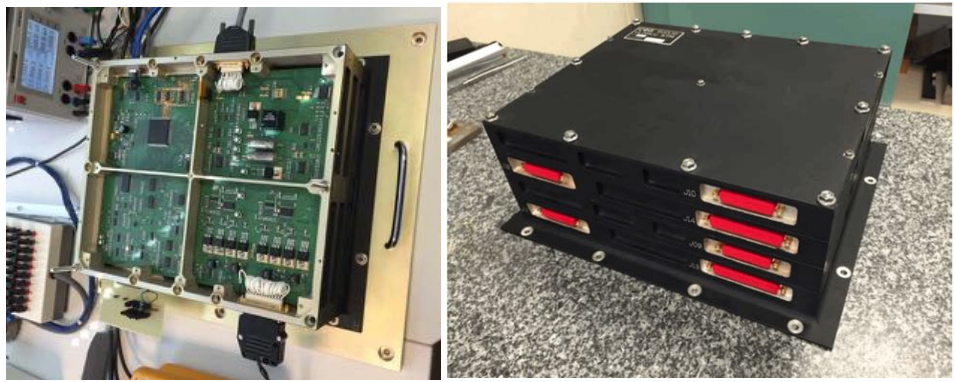

Figure 10: Two views of the EM PMCU which controls the shutter, calibration unit and provides the thermal control of the FPA.

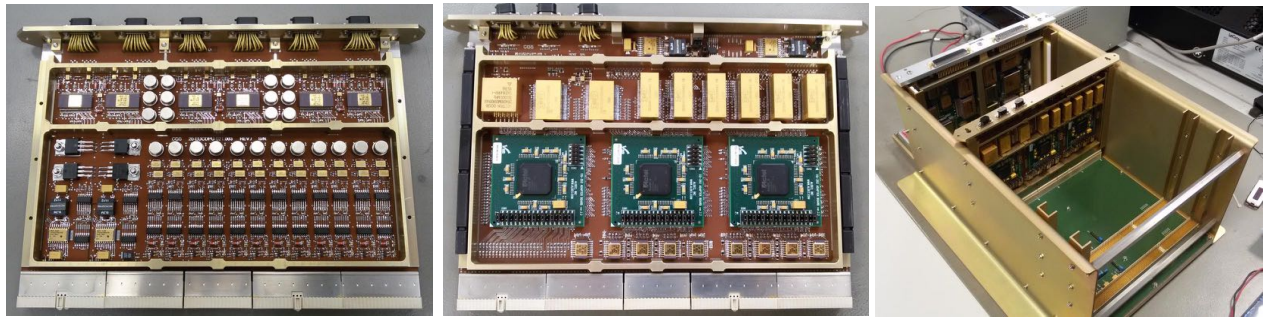

Figure 11: Two of the CDPU EM components, the multiplexer (left) and data routing (centre) boards, together with the partially populated EM CDPU structure (right).

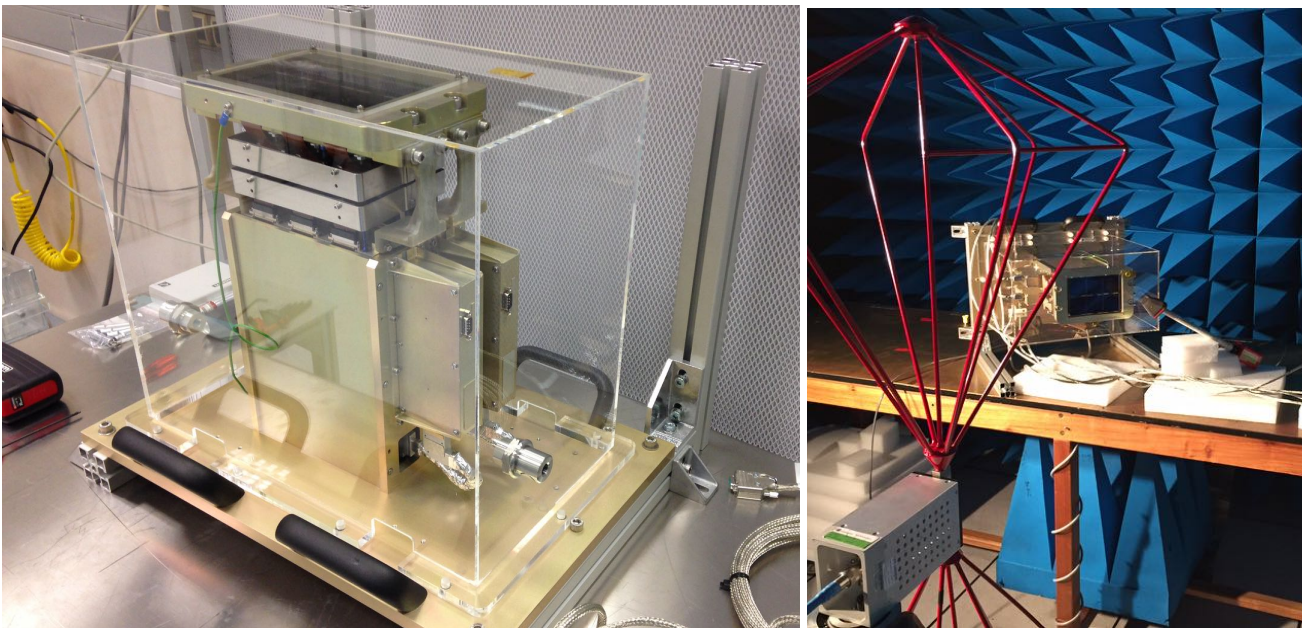

Figure 12: (left) The EM FPA comprising two half-slices each of one ROE, its R-PSU and three CCDs, making 1/6 of the full focal plane. This is fully functional as a camera with a format of 12k x 8k pixels. (right) The EM FPA undergoing radiated EMC tests.

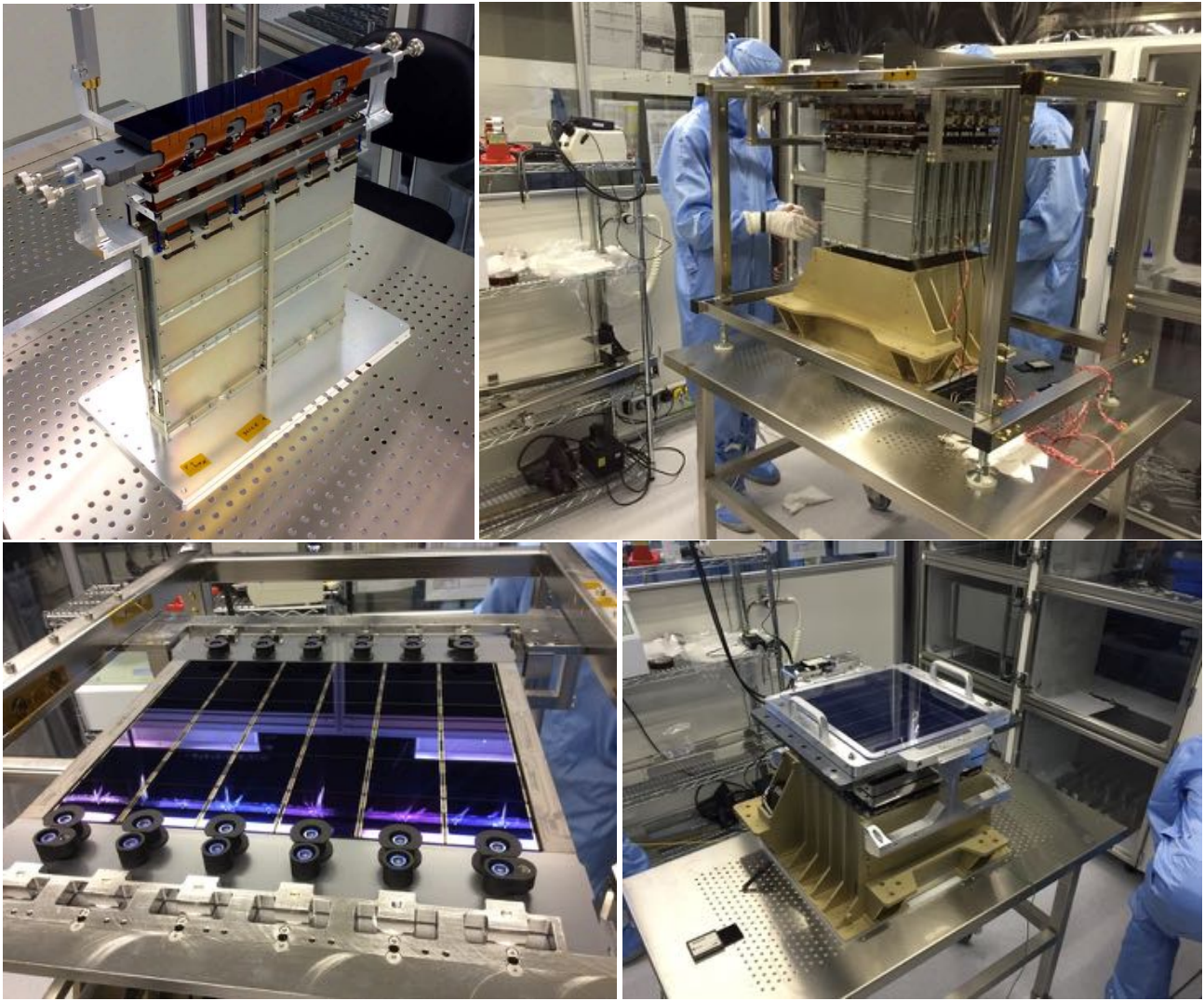

Figure 13: The assembly sequence of the STM FPA. (top left) This is a single slice of 6 CCDs on their SiC supporting bar, connected to two ROEs. The thermal shields to isolate the ambient temperature electronics from the CCDs at 153K can be seen below the row of CCDs. (top right) The six slices being lowered into the electronics structure which holds them. The ROEs and this structure form the electronics block. (bottom left) The full array of 36 CCDs made from all 6 slices integrated into the SiC structure forming the detector block. (bottom right) The full FPA excluding the R-PSUs, which will be located in the ribbed area on the side of the electronics structure. Non-flight structure fixes the detector block onto the electronics block – this will be replaced by a bracket supplied by the Payload Module when the FPA is integrated into the satellite. Also evident is the cover over the CCDs to maintain cleanliness, and, to the bottom left, contamination monitoring equipment.

Figure 14 shows bias and flat field exposures made with the flight-build CCD273-84 and EM ROE. This is part of a characterization programme to explore the detection chain characteristics in detail[21], for example the flat field colour dependence, the electrical cross-talk, linearity and the brighter-fatter effect[22]. The optimization of the performance in the presence of radiation damage is a major activity[23,24], and innovative approaches, such as tri-level clocking and slower line transfer rates have been identified and incorporated in the ROE (see Section 7).

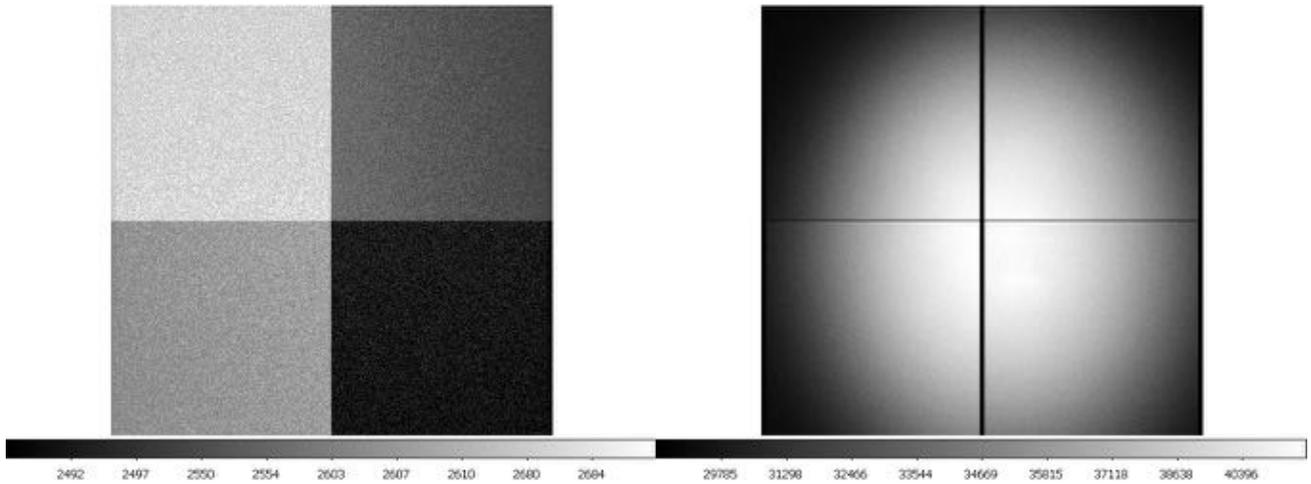

Figure 14: Bias (left) and flat field (right) exposures with the EM ROE and the EM1a CCD273-84. Slightly different bias levels are evident for each quadrant. Readout noise levels are within specification at 4.2 e⁻. In the flat field, pre- and overscan areas are evident, as well as the effect of the charge injection structure running horizontally across the CCD. The illumination is not spatially constant in this test facility.

## 6. OPERATIONS AND CALIBRATION

VIS is designed to be simple to operate because this promotes the stability needed for weak lensing measurements. Almost all of the observations are carried out with a 565 second exposure time, with the entire focal plane active. The CCDs are then read out, the data are digitised, buffered and compressed, and the process repeated. The *Euclid* Reference Survey covers 15000 deg$^2$ for the Wide Survey and ~40 deg$^2$ for the Deep Survey[25] and four separate exposures are made for each field, with displacements of 100 arcsec between them (with an additional lateral 50 arcsec for the fourth exposure). This is in order to cover gaps in the detector matrix, to permit some recovery of the spatial resolution, to minimise radiation damage impact on the data and to allow cosmic rays to be identified. This results in 50% of the sky being observed in four exposures, and another 47% with three, because of the gaps. Optimisations of the Reference Survey are currently being considered, and this may lead to a different total survey area, more exposures per field and different size displacements between them. However, any changes are required to be compatible with the current operating capabilities of the instruments and satellite.

The array takes some 80 seconds to read out in total. The VIS exposures take place during the spectroscopic exposures of NISP which have the same duration. After this, the time during the NISP photometric exposures is limited for science observations, because of pointing disturbances caused by the NISP filter wheel. However, shorter exposures will be taken during a small subset of NISP photometric exposures. These add dynamic range, improve sampling and assist in cosmic ray elimination: one or two of these short exposures can be included within the telemetry and shutter lifetime budgets. The time occupied by the NISP photometric exposures will mainly be used for VIS calibrations, such as flat and bias fields and dark exposures. Dark exposures will be of three types: in the first the dark current in the detectors (which is extremely low) can be measured, together with hot pixels (which will be relatively few given the cold operational temperature of the CCD); in the second a few lines of charge will be injected electrically into the CCD, and in the third charge will be injected over the whole region of the CCD and the CCD clocked backwards and forwards in a "pocket-pumping" mode[26]. These last two allow calibrations to be made of the radiation damage to the detector (section 7.1).

Flat field calibrations are made with the Calibration Unit. Because of possible cross-contamination with NISP, these will be done at the end of the sequence of four exposures just before a spacecraft slew. The calibration exposures are short (10 seconds), incurring no time overhead..

Linearity calibrations require a slightly different operational sequence, with different durations for each of the four exposures. In order not to inordinately constrain the shutter repeatability, the shutter will open while the CCD is being read out continuously, after which the readout will stop for a precise duration set by the synchronisation timing in VIS, before the readout begins again and the shutter closes. This will generate stellar images superimposed on illuminated strips in the image, caused by reading out when the shutter is open. These can be subtracted to produce accurate estimates of the flux in the image, and hence to determine the linearity.

Most of the calibrations will, however, be available from the science data themselves, and in particular the stars on the frame. This includes the photometric calibrations, astrometric calibrations and the calibration of the system PSF (section 7.2 below). This is highly advantageous in that these calibrations are known to be optimally representative of the instrument characteristics within each and every exposure.

## 7. PARTICULAR ISSUES

Observations from space permit the high spatial resolution which is important for weak gravitational lensing measurements of galaxies with sizes only a small fraction of an arcsec (typically ~0.3 arcsec). In addition, the absence of an atmosphere affords a high level of stability of image characteristics and a low background. It is also possible to cover the majority of the extragalactic sky. These characteristics are achievable only from space. *Euclid* VIS will detect more than 1.5 billion galaxies with a signal-noise ratio >10, and this statistical precision afforded by such a large sample is a prerequisite to constraining the nature of dark energy and dark matter. Having achieved sufficient statistical precision, the particularly difficult aspect of the weak lensing measurements then concerns the control of the statistical biases. This sets exceptionally tight requirements[9,27] on the knowledge of the instrumental characteristics, and particularly on the knowledge of the shape of the PSF. Two major issues affect this knowledge. We also mention a third which is currently the focus in several other experiments.

### 7.1 Radiation

Radiation damage effects in electronics and detectors at the orbit of *Euclid* near the L2 point are largely the result of Solar ions with ~MeV energies. These cause lattice damage in the Si within the pixels and readout registers, creating traps for electrons which impact the transfer of charge from pixel to pixel when the CCD is read out. This leaves a charge trail behind each image as it is moved down the CCD column, and then another in the orthogonal direction as it is transferred along the readout register (Figure 15). The radiation damage effect is parameterised initially through the Charge Transfer Inefficiency (CTI), which measures the fraction of charge lost after the pixel charge contents have been clocked to the readout register, but in *Euclid* a much more detailed characterisation is required in which the details of the charge trail are known to high accuracy.

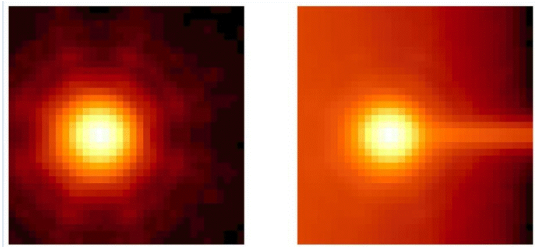

Figure 15: The effect of radiation damage in the CCD. The figure on the left is a simulated source through the *Euclid* system without radiation damage (O. Boulade, CEA Saclay using code from A. Short, ESA) and the right shows the image with radiation damage and cosmic background included when the CCD is read out to the left side of the image. Both are on a log intensity scale to show low-level effects.

These effects are particularly relevant to *Euclid* VIS, because they change the shape of the image, directly affecting the primary quantity to be measured. The traps in the lattice can be filled at least partially with the charge injection lines (section 6) to ameliorate their effect. The same charge injection lines can be used to infer the time constants and overall densities of the traps. Moreover, the location and time constants of the traps can be identified using the pocket pumping technique of shifting charge backwards and forwards within the CCD, once charge injection lines have been injected throughout the device (section 6). Phenomenological[28,29] and Monte-Carlo models are used to predict the effects of the trapping, and to reverse the charge trailing this causes (see section 8).

### 7.2 The VIS point spread function

The PSF is composed of the optical PSF, the spacecraft pointing jitter and the detector internal charge spread convolved together, and sampled on the CCD pixel grid. The PSF is slightly undersampled by this grid, and some recovery of the spatial frequencies will rely on the combination of the three to four individual exposures.

The main contributor to the PSF is the optical system. Given that it is diffraction-limited, the core of the PSF scales nearly inversely with wavelength. There will be some ellipticity in the images resulting from the optical aberrations, and imperfections in manufacture and alignment of the elements. The broadening of the PSF from the spacecraft pointing jitter during an exposure will be different from exposure to exposure, and will depend on the characteristics of the attitude control system. The internal charge spread within the CCDs will depend on their particular operation, and in particular the number of phases held high during the integration. These define the shape of the electrical potential in the column direction, while in the row direction the potential is defined by the column stops. Besides this is the residual effect of the radiation-induced charge trailing (on the order of a few $\times 10^{-3}$ of the uncorrected trailing), after it has been corrected in the data processing[28,29]. The shape of the detector-induced contribution to the PSF is therefore slightly non-axisymmetric, which adds ellipticity aligned with the pixel grid to the measured galaxy shapes.

The degree to which the model of the PSF is a faithful representation of the real PSF must be extremely carefully controlled through a detailed budget breakdown[9] which includes all of the instrumental and calibration effects. These must not exceed the additive and multiplicative biases permitted in Table 1. A model PSF can be constructed by the superposition of different PSF components, and principal components analysis (PCA) was initially used to generate the components. This has been superseded in the ground segment development for the weak lensing with more sophisticated Fourier techniques. Because the major contribution to the PSF is diffraction, generating the PSF requires the spectral energy distribution of the stars, and of the galaxies to be known. This will be available from the combination of ground-based, VIS and NISP photometry. The uncertainties and biases within this diverse photometric resource will require careful analysis and modelling in order to keep their impact within allocation. In addition, the colour dependence of the flat field, and hence in the residual corrected flat field maps, may require a spectral energy distribution-dependent flat field to be generated for each bright star, produced from the 6 bandpasses provided by the Calibration Unit.

### 7.3 Brighter-fatter Effect and Tree Rings

The PSFs of some imaging systems using thick high-resistivity CCDs have been found to be intensity dependent, in the sense that PSFs with more photo-electrons are less peaked than those with fewer. This is known as the "brighter-fatter effect". In VIS, for the CCD273-84 with 40μm thick 1500 Ωcm material, the effect is less pronounced[22], but will still require attention. Pending further investigation, the effect may be treated as an additional convolution, an adjustment of the weight function, or the use of higher-order moments.

Furthermore, a concentric pattern, termed "tree rings", is seen in flat fields taken using thicker CCD devices. These cause astrometric displacements within the image, and perhaps photometric variations. "Tree rings" are evident in summations of many flat fields in the CCD273-84, but they are at an extremely low level. The degree of mitigation required in the science ground system processing depends on the spatial frequency of the effect, and whether it will be recovered from the astrometric calibration is being considered.

## 8. PERFORMANCE

Current best estimates with radiometric modelling predicts that VIS will perform better than the $m_{AB} \geq 24.5$ requirement, reaching $m_{AB} = 24.9$ (10σ) for a 0.43 arcsec extended source in a 1.3 arcsec diameter aperture in its $r+i+z$ band from three exposures of 565 sec. This assumes end-of-life conditions and moderately high levels of Zodiacal background. Half of the sources will be covered by four exposures from the dither pattern. Detector noise (mostly readout noise) will be <4.5e⁻. Stellar sources reach 10σ approximately 0.1 mag fainter, and saturate at $R$=18.3. End-to-end simulations predict that the system PSF will have a full-width-half-maximum 0.17 arcsec. The system ellipticity is ~2%. These indicate that the size and depth of the survey will be sufficient to produce the large number of galaxies required to constrain the cosmological parameters. The largest task in assessing the performance is to quantify the contributions to the biases in the weak lensing measurements, particularly requiring the high level of knowledge of the PSF.

It is clear that radiation damage will be a critical issue for VIS, and a substantial programme is in place to understand and quantify its effect[23,24], and to introduce ameliorating measures into the CCD development programme, where these might be possible. CTI correction is limited by the presence of readout noise in the CCD and ROE[28]; a reduction in this has already been achieved. Continuing measures noted above include radiation testing and characterisation, development of software models of varying complexity and of software algorithms for the final data analysis[24,28,29]. Many parameters need to be determined for inclusion in the models: in particular the time constants and cross sections of the different trap species, and the beneficial impacts of a non-zero optical background (mostly arising from Zodiacal light) and charge injection have been quantified. The multi-exposure strategy (dithering) in section 6 above is an important ingredient in isolating the effect, as each source will be exposed at different locations on the CCD with consequent trails of different extents. The calibration strategy (section 6 again) is designed to provide the information required for the models. Full end-to-end performance predictions of continuously increasing fidelity continue to be developed to quantify its impact. It is important to note however that the effects manifest themselves largely on the angular scale of a CCD quadrant (as there are four readout nodes in the CCD273-84), and not over the full range of scales measured in the 2-point statistics used to infer the cosmological parameters. The effect of this has been investigated[30]. The residuals after modelling are all located in the detector (rather than the cosmic) reference frame, and their contribution quantified by stacking images. Differences in the amplitude of the residuals from detector to detector, and in the relative spatial phasing of the pattern given the overlap of exposures, affects the final impact of these, and also the extent of the mixing of angular scales in the power spectrum. More broadly, however, the impact is scale dependent, which provides margin against the scale-independent allocation for the shear systematic biases in Table 1.

Once the images have been corrected for the charge trailing caused by radiation damage, and other more standard calibrations have been applied, the image is available both for legacy science use by the community at large, and for the weak lensing analysis. This analysis requires the effective PSF to be known for each galaxy as described in section 7.2 above. The galaxy ellipticities can then be measured, taking into account this known, locally derived PSF used in the measurement, and galaxy colour gradients as a function of redshift, from a calibration sample. As currently evaluated, and taking all significant contributions into account, the PSF modelling can be carried out sufficiently accurately to limit the biases to within the allocated limits at the bottom of Table 1 for each image independently. No PSF stability need be assumed from one image to the next, and hence any stability from one exposure to the next provides margin in reaching the required low bias levels. In practice, significant gains will be made if the stability of the *Euclid* optical system is as expected. The aggregate of the measured galaxy ellipticities on different spatial scales constrains the weak lensing and hence the cosmological parameters such as the equation of state of dark energy to the required levels[9].

## 9. SUMMARY

The *Euclid* VIS instrument, together with the *Euclid* optical system and survey and calibration strategies will produce high spatial resolution images of most of the extragalactic sky, and through measurements of the ellipticities of over $10^9$ galaxies and a rigorous control of systematic biases, will provide tight constraints on cosmological models. The instrument design is stable and through early development of critical subsystems, programmatic and technical risks have been minimised. The projected instrument performance has been analysed comprehensively relative to a detailed budget allocation to all of the salient effects: this analysis in itself has required detailed understanding. Current evaluations indicate that all of the science requirements can be met.